\documentclass[twocolumn]{aastex61}

\newcommand\aastex{AAS\TeX}

\shorttitle{\aastex\ Launching PNS kick with asymmetric neutrino emissions}
\shortauthors{Nagakura et al.}

\begin{document}
\title{Possible early linear acceleration of proto-neutron stars via asymmetric neutrino emission in core-collapse supernovae}
\correspondingauthor{Hiroki Nagakura}
\email{hirokin@astro.princeton.edu}

\author{Hiroki Nagakura}
\affiliation{Department of Astrophysical Sciences, Princeton University, Princeton, NJ 08544, USA}

\author{Kohsuke Sumiyoshi}
\affiliation{Numazu College of Technology, Ooka 3600, Numazu, Shizuoka 410-8501, Japan}

\author{Shoichi Yamada}
\affiliation{Advanced Research Institute for Science \&
Engineering, Waseda University, 3-4-1 Okubo,
Shinjuku, Tokyo 169-8555, Japan}
\affiliation{Department of Science and Engineering, Waseda University, 3-4-1 Okubo, Shinjuku, Tokyo 169-8555, Japan}

\begin{abstract}
In this {\it Letter}, we present the result of an axisymmetric CCSN simulation conducted with appropriate treatments of neutrino transport and proper motions of proto-neutron star (PNS), in which a remarkable PNS acceleration is observed in association with asymmetric neutrino emissions by the time of $300$ms after bounce. We find that these asymmetric neutrino emissions play important roles in the acceleration of PNS in this phase. The correlation between the PNS proper motion and the asymmetric ejecta is similar to that in the neutron star (NS) kick of hydrodynamic origin. Both electron-type neutrinos ($\nu_{\rm e}$) and their anti-particles ($\bar{\nu}_{\rm e}$) have $\sim 10\%$ level of asymmetry between the northern and southern hemispheres, while other heavy-leptonic neutrinos ($\nu_x$) have much smaller asymmetry of $\sim 1 \%$. The emissions of $\bar{\nu}_{\rm e}$ and $\nu_x$ are higher in the hemisphere of stronger shock expansion whereas the $\nu_{\rm e}$ emission is enhanced in the opposite hemisphere: in total the neutrinos carry some linear momentum to the hemisphere of the stronger shock expansion. The asymmetry is attributed to the non-spherical distribution of electron-fraction ($Y_e$) in the envelope of PNS. Although it is similar to LESA (lepton-emission self-sustained asymmetry), the $Y_e$ asymmetry seems to be associated with the PNS motion: the latter triggers lateral circular motions in the envelope of PNS by breaking symmetry of the matter distribution there, which are then sustained by a combination of convection, lateral neutrino diffusion and matter-pressure gradient. Our findings may have an influence on the current theories on the NS kick mechanism although long-term simulations are required to assess their impact on the later evolution.
\end{abstract}

\keywords{supernovae: general---neutrinos---radiative transfer---hydrodynamics}



\section{Introduction} \label{sec:intro}
There is growing evidence that core-collapse supernovae (CCSN) are highly non-spherical phenomena. The role of the asymmetry during the development of explosion is certainly important: for instance, it is well established that multi-dimensional (multi-D) hydrodynamic instabilities are one of the key ingredients to trigger explosion (see recent reviews, e.g., by \citet{2015PASA...32....9F,2016NCimR..39....1M,Janka:2016fox,2016PASA...33...48M}). As a consequence of the non-spherical explosions, the nascent neutron star (NS), which is left behind at the center of explosion, inevitably receives a kick (see, e.g., a review by \cite{2001LNP...578..424L} and references therein). It may account for the fast proper motions of pulsars, which move typically at a few hundreds km/s \citep{1994Natur.369..127L,2002ApJ...568..289A,2005MNRAS.360..974H,2006ApJ...643..332F,2005ApJ...630L..61C,2007ApJ...670..635W}.

The NS kick mechanism has been one of the primary research subjects in the CCSN community. If it is of hydrodynamic origin, in which case NS is accelerated by hydrodynamic or gravitational forces or both exerted by the asymmetric ejecta, it should be kicked in the opposite direction to the hemisphere of stronger explosion. Recently, X-ray observations of young CCSN remnants by \citet{2017ApJ...844...84H,2019arXiv190406357H,2018ApJ...856...18K} found such a correlation between the kick direction and the asymmetry of ejecta, and hence supported the hydrodynamic mechanism. However, such a correlation is just one of the necessary conditions and can not be a smoking gun of that particular mechanism, since other mechanisms may generate the same correlation. On the other hand, previous studies suggest that asymmetric neutrino emissions from a proto-neutron star (PNS) are not large enough to accelerate PNS to a few hundreds km/s \citep{2004PhRvL..92a1103S,2006A&A...457..963S,2010PhRvD..82j3016N,2012MNRAS.423.1805N,2013A&A...552A.126W,2017ApJ...837...84J,2018ApJ...865...61G,2018arXiv181105483M}. Other mechanisms also require rather extreme conditions to produce the large PNS kick as observed. From these facts the hydrodynamic origin is currently the most favored mechanism for the NS kick. Note that in this mechanism the NS kick is produced at rather late times, when the stalled shock has been already revived and an explosion has been firmly established.

In this {\it Letter}, we study a possible initiation of PNS acceleration in the earlier phase based on the result of one of our latest axisymmetric CCSN simulations. This study is motivated by the fact that there has been no CCSN simulation that handles the PNS proper motions and their feedback to both hydrodynamics and neutrino transport fully self-consistently. Note that, although \citet{2006A&A...457..963S} also implemented a moving mesh technique to treat PNS proper motions, it was applied to the hydrodynamics but not to the neutrino transport. In addition, they excised the interior of PNS in their simulations, which is a clear distinction from the simulation presented in this paper. This is hence the first-ever attempt to do that in Boltzmann neutrino-radiation-hydrodynamic simulations. We find that the self-consistent treatment turns out to be crucial: indeed, a remarkable PNS motion occurs in association with large asymmetric neutrino emissions

The CCSN dynamics obtained in our simulation is aligned with the neutrino-heating explosion mechanism although the computation is not long enough to make a firm conclusion. Since we did not incorporate magnetic fields, rotations and non-standard neutrino physics in this simulation, the result presented in this paper is distinct from other mechanisms to produce PNS proper motions that advocate asymmetric neutrino emissions \citep{1993A&AT....3..287B,2003PhRvD..68j3002F,2008PhRvD..77l3009K,2008A&A...489..281S}. Interestingly, although the property of asymmetric neutrino emissions is the same as in LESA (lepton-emission self-sustained asymmetry), the origin may be different from those discussed in \citet{2014ApJ...792...96T,2018arXiv180910150G,2018arXiv181205738P}. We find in fact that the non-spherical neutrino emissions are caused by the asymmetry in the electron fraction ($Y_e$) distribution in the envelope of PNS, which is in turn associated with the PNS motion.

\section{Method and Model} \label{sec:model}
We carry out an axisymmetric simulation of CCSN, solving Boltzmann equations for multi-species, multi-energy and multi-angle neutrino transport. The special relativistic effect in neutrino transport is fully taken into account \citep{2014ApJS..214...16N}. A moving mesh technique is implemented in order to treat PNS proper motions self-consistently \citep{2017ApJS..229...42N}, which is the most important feature for this paper. The reliability of our code was established by a detailed comparison with a Monte Carlo simulation \citep{2017ApJ...847..133R}. Very recently, we further developed a new treatment of the momentum feedback from neutrino to matter \citep{2019arXiv190610143N}. In this method, the momentum exchange is calculated with the energy-momentum tensor instead of by the direct integral of collision term in the diffusion regime, which substantially improved the accuracy of momentum conservation in our simulations.

Most parts of the numerical setup and input physics are the same as those used in \citet{2018ApJ...854..136N}. We solve radiation-hydrodynamics with Newtonian self-gravity consistently. Note that we recently improved the treatment of nuclear weak interactions, which is meant to be used with the multi-nuclear EOS based on the variational method \citep{2013NuPhA.902...53T,2017JPhG...44i4001F}: we constructed new tables of electron captures by heavy and light nuclei and positron captures by light nuclei, employing the nuclear abundances provided by the EOS (see \citet{2019ApJS..240...38N} for more details). We incorporate all these updates in the simulation.

A 11.2 $M_{\sun}$ progenitor model by \citet{2002RvMP...74.1015W} is used. The spherical coordinates $(r, \theta)$ are meshed with $384(r) \times 128(\theta)$ grid points covering $0 \le r \le 5000{\rm km}$ and $0^{\circ} \le \theta \le 180^{\circ}$ in the meridian section, respectively. The momentum space for neutrinos with the energy range of $0 \le \varepsilon \le 300{\rm MeV}$ and the entire solid angle ($\Omega$) is covered with $20(\varepsilon) \times 10(\tilde{\theta}) \times 6(\tilde{\phi})$ grid points, respectively, where $\tilde{\theta}$ and $\tilde{\phi}$ are the zenith and azimuth angles of neutrino momentum. We consider three neutrino species: electron-type neutrinos $\nu_{\rm e}$, electron-type anti-neutrinos $\bar{\nu}_{\rm e}$ and all the others collectively denoted by $\nu_x$. The simulation is run up to $300$ms after bounce.

\section{CCSN dynamics and PNS kick} \label{sec:PNSKick}

\begin{figure}
\vspace{15mm}
\epsscale{1.3}
\plotone{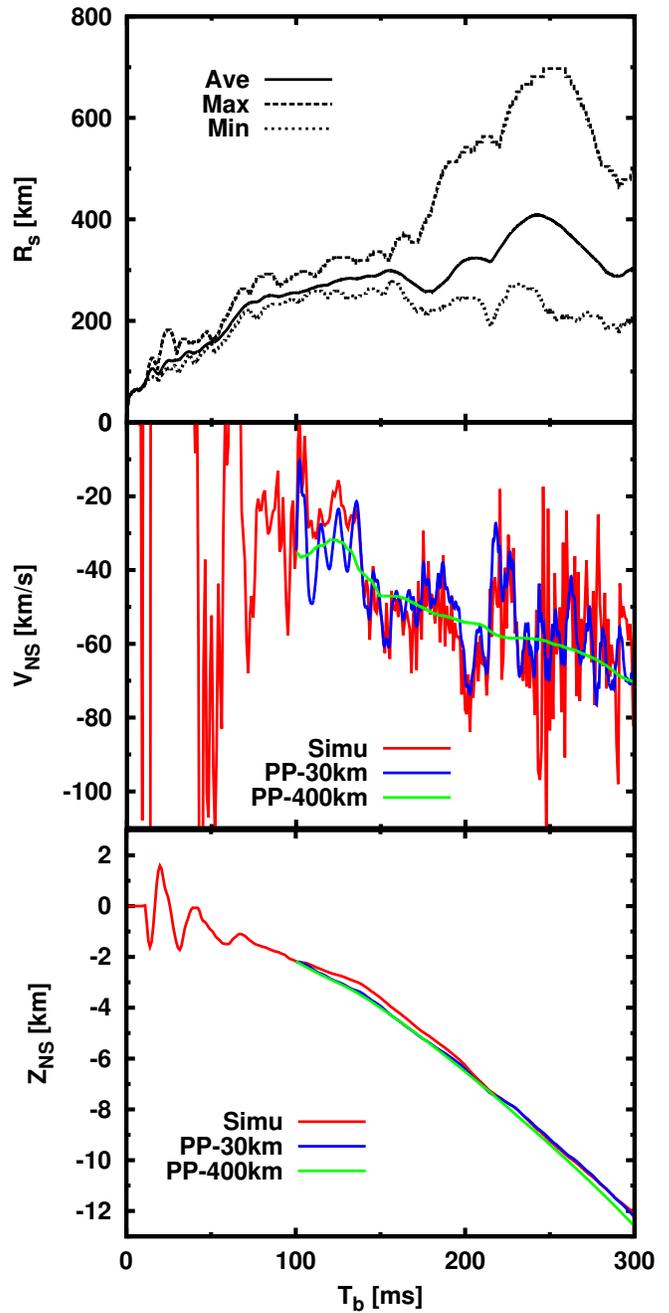}
\caption{Top: time evolutions of the maximum, minimum and average shock radii. Middle: time evolution of the PNS velocity in the simulation (red line). The blue and green lines show the post-processed (PP) results in the PNS kick analysis (see the text for details). Bottom: time evolution of the spatial displacement of PNS (red line) in the z direction. The blue and green lines display the results obtained in the post-processing.
\label{graph_tevo_rsh_VNS_ZNS}}
\end{figure}

Prompt convection begins at ${\rm T_b} \sim 10$ms (${\rm T_b}$ denotes the time after bounce) and generates non-spherical fluctuations that will become a seed perturbation for later fluid instabilities in the post-shock flow. The neutrino-driven convection starts to appear at ${\rm T_b} \sim 100$ms and triggers a rapid shock expansion in the northern hemisphere at ${\rm T_b} \sim 180$ms (see the top panel in Fig.~\ref{graph_tevo_rsh_VNS_ZNS}). Note that we can not clearly judge at ${\rm T_b} = 300$ms (the end of our simulation) if this model leads to explosion or not (see also \citet{2006A&A...457..281B,2012ApJ...756...84M,2014ApJ...786...83T,2019arXiv190109055P}), and a longer computation is definitely required for that purpose. Nonetheless we find a coherent PNS motion clearly up to this point, which may lead to a launch of the NS kick if this model really explodes and, moreover, if it survives all that will happen afterward. Since we expect that this proper motion will continue in the subsequent phase of a runaway asymmetric shock expansion, it is well worth a detailed analysis.

The red lines in the middle and bottom panels of Fig.~\ref{graph_tevo_rsh_VNS_ZNS} display the time evolutions of the velocity ($V_{\rm NS}$) and spatial displacement ($Z_{\rm NS}$) of PNS, respectively\footnote{The velocity actually corresponds to the shift vector, which is chosen to track approximately the motion of PNS (see \citet{2017ApJS..229...42N}).}. Since the large temporal variations at ${\rm T_b} \lesssim 100$ms due to the prompt convection are rather random and will not be directly relevant to the PNS acceleration, we will focus on the later phase ${\rm T_b} > 100$ms hereafter. Although $V_{\rm NS}$ is fluctuating rapidly because of our way of implementation of the moving mesh technique\footnote{See \citet{2017ApJS..229...42N} for more details on the origin of the noise.}, it is clear that the PNS receives a linear momentum to the south from ${\rm T_b} \sim 150$ms. At the end of the simulation, the velocity and the displacement reach $V_{\rm NS} \sim 60$km/s and $Z_{\rm NS} \sim 12$km, respectively, and the PNS is still accelerating in the same direction.

\begin{figure}
\vspace{15mm}
\epsscale{1.2}
\plotone{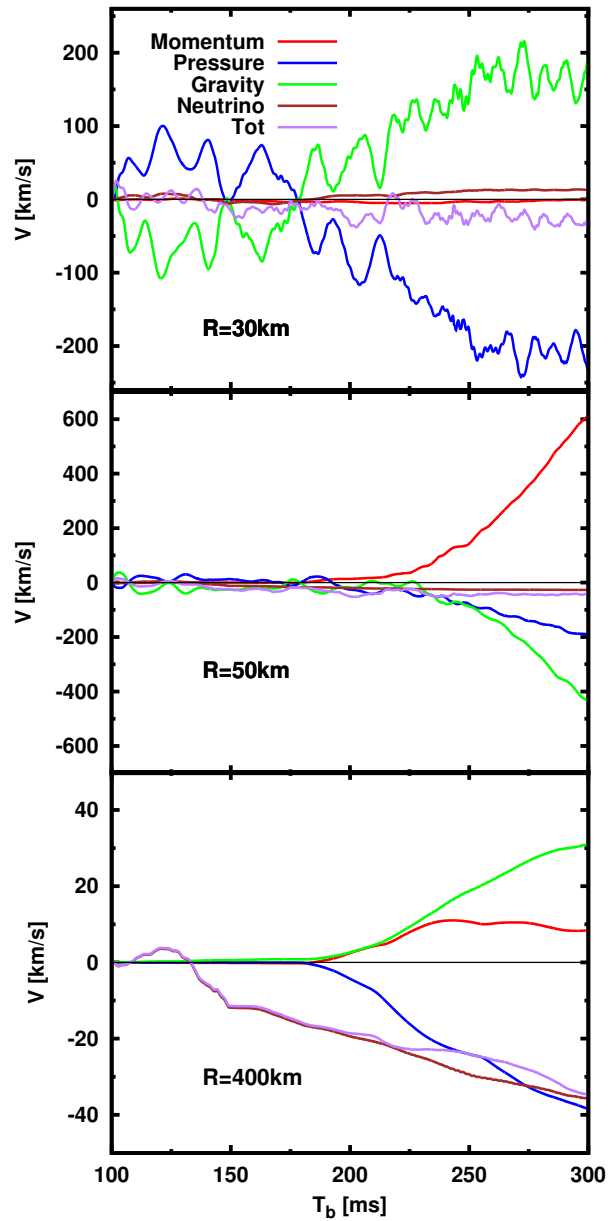}
\caption{Time evolutions of various contributions to the PNS kick. From top to bottom, we display the results of individual terms in Eqs.~(\ref{eq:ac_tot}) and (\ref{eq:ac_indivi}) at $R = 30, 50$ and $400$km, respectively.
\label{graph_tevo_decomposeNSkick}}
\end{figure}

Following the common practice, we post-process the result to see what forces dictate the PNS motion. We evaluate the matter acceleration ($a_{\rm tot}$) in the region of $r \le R$, where $r$ denotes the radius measured from the mass center of PNS, which can be written as:
\begin{eqnarray}
a_{\rm tot}(R) = a_{\rm m}(R) + a_{\rm p}(R) + a_{\rm g}(R) + a_{\nu}(R), \label{eq:ac_tot}
\end{eqnarray}
where $a_{\rm m}, a_{\rm p}, a_{\rm g}$ and $a_{\nu}$ represent the contributions from the momentum flux and matter pressure, gravity and neutrino-matter interactions, respectively. Each term can be expressed as:
\begin{eqnarray}
&& a_{\rm m} (R) \equiv - \frac{ R^2 \int_{{r = R}} \rho v_{z} v^{r} d\Omega }{M(R)}, \nonumber \\
&& a_{\rm p} (R) \equiv - \frac{ R^2 \int_{{r = R}} P {\rm cos}{\theta} d\Omega }{M(R)}, \nonumber \\
&& a_{\rm g} (R) \equiv - \frac{ \int_{{r \le R}} \rho \psi_{,z} dV }{M(R)}, \nonumber \\
&& a_{\rm \nu} (R) \equiv - \frac{ \int_{{r \le R}} G_{z} dV }{M(R)}, \label{eq:ac_indivi}
\end{eqnarray}
where $\rho, v_{z}, P, \psi$, $G_{z}$ and $V$ denote the baryon mass density, the z-component of fluid velocity, the matter pressure, the gravitational potential, the z-component of momentum transfer from neutrino to matter\footnote{The exact form of neutrino-matter interaction can be seen in (\citet{2017ApJS..229...42N} and Nagakura et al. 2019 in prep)} and the spatial volume, respectively. $M(R)$ is the baryon mass in the region ($r \le R$). We display the result in Fig.~\ref{graph_tevo_decomposeNSkick}, which is obtained by integrating Eqs.~(\ref{eq:ac_tot}) and (\ref{eq:ac_indivi}) at three different radii ($R = 30, 50$ and $400$km). Note that the time-integration is started from ${\rm T_b} = 100$ms to remove the contribution from the prompt convection.

At $R=30$km (see the top panel of Fig.~\ref{graph_tevo_decomposeNSkick}), the anisotropies of gravity and matter pressure are the dominant forces but almost cancel out each other. The neutrino contribution is roughly $10 \%$ of them, and the anisotropy in momentum flux is negligible. Note that the blue lines in the middle and bottom panels of Fig.~\ref{graph_tevo_rsh_VNS_ZNS} are calculated from $a_{\rm tot}$ at $R=30$km. We find that the velocities and displacements obtained in the post-process are consistent with the PNS proper motion observed in the simulation\footnote{Note that the initial values of $V_{\rm NS}$ and $Z_{\rm NS}$ for the post-process are derived from the simulation.}. At first glance, the anisotropic matter pressure may look dominant as a driving force of PNS motion (see the phase ${\rm T_b} \gtrsim 150$ms). It is too early to reach the conclusion, since the condition strongly depends on the radius, however. In fact, at $R=50$km (in the middle panel of Fig.~\ref{graph_tevo_decomposeNSkick}), the gravity term changes sign and overwhelms the pressure term. However, the sum of gravity and pressure contributions is almost canceled by that from the momentum-flux. On the other hand, the neutrino contribution, which has also opposite signs at these two radii, is comparable to the total acceleration. This is an indication that neutrinos are not streaming completely freely but are still mildly coupled with matter in the region of $30 \lesssim r \lesssim 50$km. Because of this strong dependence on radius, it is hard to identify which component is the most dominant player to drive the PNS proper motion (see also \citet{2010PhRvD..82j3016N}).

On the other hand, we find another interesting result from the same analysis but applied to a larger radius ($R=400$km), which is shown in the bottom panel of Fig.~\ref{graph_tevo_decomposeNSkick}. The contributions from gravity and hydrodynamic forces become much smaller than those at the smaller radii, and the neutrino contribution overwhelms other ones. This result is not surprising, since all acceleration terms except for the neutrino term tend to zero asymptotically. What is interesting here is that $a_{\rm tot}(R=400{\rm km})$ is comparable with the acceleration of the PNS obtained in the simulation. As a matter of fact, $V_{\rm NS}$ and $Z_{\rm NS}$ calculated in the post-process are roughly consistent with the PNS motion observed in the simulation (see green lines in the middle and bottom panels of Fig.~\ref{graph_tevo_rsh_VNS_ZNS}). This fact indicates that the role of neutrinos cannot be ignored in the PNS acceleration.

It is important to emphasize that this result is brand new in the CCSN simulation, since self-consistent handling of the feedback from PNS proper motions to both hydrodynamics and neutrino transport is indispensable but was not realized one way or another in previous studies. In fact, most of previous simulations employed some pragmatic prescriptions to either PNS proper motions or neutrino transport or both and they may have artificially suppressed the phenomenon. More detailed studies are needed in the future to confirm it, though.

\begin{figure}
\vspace{15mm}
\epsscale{1.2}
\plotone{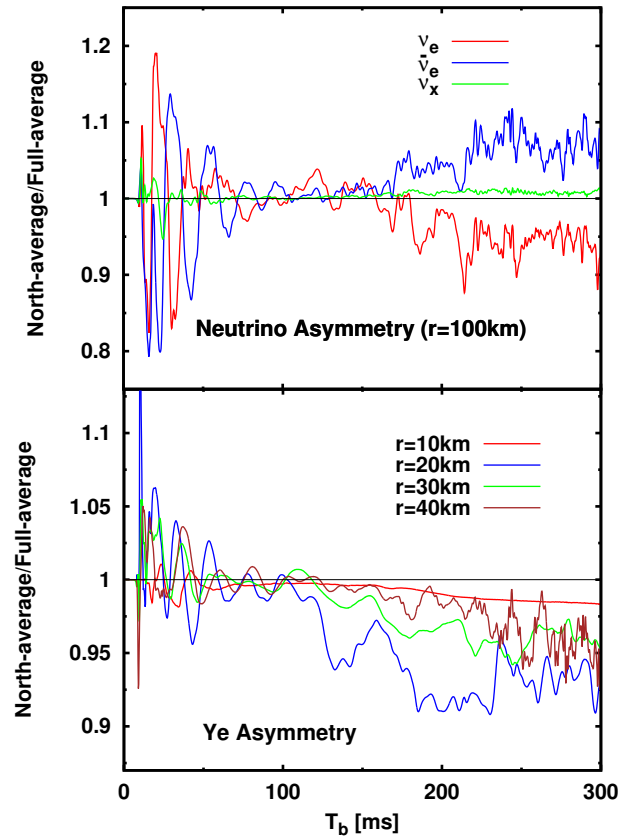}
\caption{Top: time evolutions of the neutrino asymmetry at $r=100$km. It is measured by the ratio of the northern-hemispheric average to the full-spherical average of neutrino luminosity. By definition, it is larger than 1 if the northern hemisphere has a higher luminosity than the southern hemisphere and vice versa. Bottom: same as the top panel but for $Y_e$ asymmetry. The color represents the results at different radii.
\label{graph_tevo_AsymDeg_Neutrino_Ye}}
\end{figure}

\begin{figure}
\vspace{15mm}
\epsscale{1.2}
\plotone{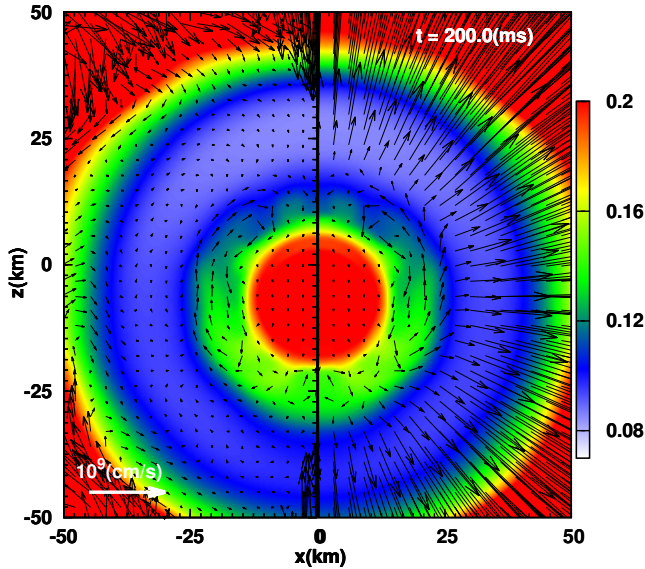}
\caption{The $Y_e$ distributions in color with the matter velocity vector (left) and the energy flux of $\nu_{\rm e}$ measured in the laboratory frame (right) (normalized by the energy density) as vectors at ${\rm T_b} = 200$ms.
\label{g2Dvec_Yedis_V_FovNnueLb_Paper200ms}}
\end{figure}

\section{Asymmetric neutrino emissions} \label{sec:Asymneutri}
We turn our attention to the asymmetric neutrino emission. As shown in the top panel of Fig.~\ref{graph_tevo_AsymDeg_Neutrino_Ye}, coherent asymmetric neutrino emissions are clearly seen from ${\rm T_b} \sim 150$ms, which roughly coincides with the time of the initiation of the asymmetric shock expansion and the PNS acceleration. The $\nu_{\rm e}$ luminosity is stronger in the southern hemisphere (the direction of the PNS motion) whereas the $\bar{\nu}_{\rm e}$ luminosity has the opposite trend (higher in the direction of the stronger shock expansion). Although the asymmetry in $\nu_x$ is smaller than in others, it is higher in the northern hemisphere.

Both asymmetries in $\nu_{\rm e}$ and $\bar{\nu}_{\rm e}$ emissions are due to the asymmetric distribution of $Y_e$ around the surface of the PNS (see the bottom panel of Fig.~\ref{graph_tevo_AsymDeg_Neutrino_Ye} and Fig.~\ref{g2Dvec_Yedis_V_FovNnueLb_Paper200ms}). The $Y_e$ in the northern hemisphere is somewhat smaller than in the southern hemisphere: for instance, it is $\sim 10 \%$ smaller than the angle average for $r = 20$km at ${\rm T_b} \gtrsim 200$ms. Since the high-$Y_e$ environment produces stronger $\nu_{\rm e}$ and weaker $\bar{\nu}_{\rm e}$ emissions, this is consistent with the trend of the neutrino asymmetry in the simulation. Note that this region is semi-transparent to neutrinos: in fact $\nu_{\rm e}$ and $\bar{\nu}_{\rm e}$ start to decouple with matter in the region of $20 \lesssim r \lesssim 25$km at ${\rm T_b} = 200$ms, which can be understood from a comparison between the vector fields in the left and right panels in Fig.~\ref{g2Dvec_Yedis_V_FovNnueLb_Paper200ms}. At $r \sim 25$km\footnote{The matter density is $\sim 7 \times 10^{12} {\rm g/cm^3}$ at this radius.} the fluid velocity (displayed on the left panel) and the $\nu_{\rm e}$ energy flux (displayed on the right one) deviate from each other.

On the other hand, it is not so easy to identify the cause of the asymmetry in the $\nu_x$ luminosity as the asymmetry is subtle. It may be attributed to the fact that the mass accretion in the northern hemisphere is smaller than in the southern hemisphere, which is caused by the asymmetric shock expansion. As discussed in \citet{2019ApJS..240...38N}, the weaker mass accretion facilitates the $\nu_x$ diffusion from PNS, and may be responsible for the higher $\nu_x$ luminosity in the northern-hemisphere. It is noted that the non-spherical mass accretion may also contribute to the asymmetries in $\nu_{\rm e}$ and $\bar{\nu}_{\rm e}$ luminosities. The effect, however, seems to be minor compared with the contribution from the asymmetric $Y_e$ distribution in the PNS envelope (see below), since the former asymmetry is subtler than the latter.

The asymmetries in $\nu_{\rm e}$ and $\bar{\nu}_{\rm e}$ luminosities reach $\sim 10 \%$. We find that the linear momentum transfer from $\bar{\nu}_{\rm e}$ to matter is slightly larger than that from $\nu_{\rm e}$, which induces the PNS acceleration in the opposite direction to the stronger shock expansion. The asymmetric $\nu_x$ emission also enhances the acceleration of PNS.

Note that the sense of the correlation between the asymmetries in $\nu_{\rm e}$ and $\bar{\nu}_{\rm e}$ emissions and that in the shock expansion is the same as in LESA. However, there is no clear indication of deflected accretion streams in the post-shock flow, which were originally suggested as a possible cause of LESA by \citet{2014ApJ...792...96T}. Instead, the asymmetry of the $Y_e$ distribution is sustained by a long-lived coherent lateral circular fluid motion in the envelope of PNS. In fact, such a coherent motion emerges soon after the asymmetric shock expansion occurs. It is indeed recognized at $10 \lesssim r \lesssim 25$km between the equator and the north-pole (see the vector field of fluid velocity displayed in the left panel).

\begin{figure}
\vspace{15mm}
\epsscale{1.2}
\plotone{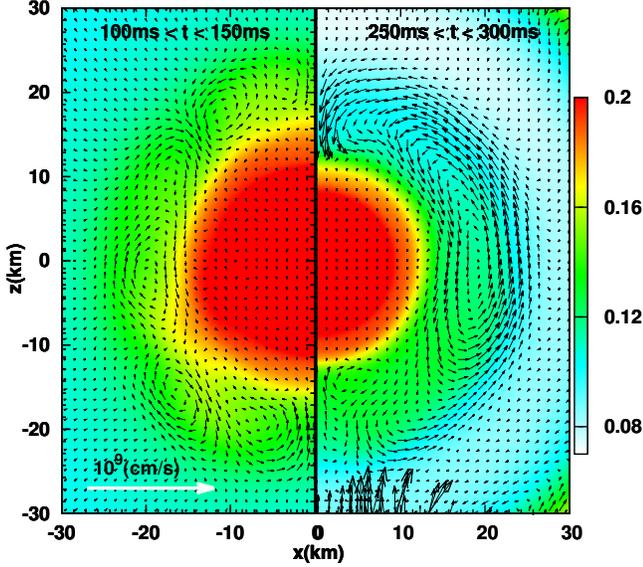}
\caption{Time average of $Y_e$ in color and fluid velocities as vectors. The left and right panels show the results for the averages over $100{\rm ms}<t<150$ms and $250{\rm ms}<t<300$ms, respectively. Note that the spatial scale and color coding are different from those in Fig.\ref{g2Dvec_Yedis_V_FovNnueLb_Paper200ms} and the coordinate origin ((x,z)=(0,0)) in this figure is shifted to the mass center of PNS. See the text for more details.
\label{g2Dvec_Yedis_V_contrast}}
\end{figure}

It is a bit obscured, however, by short-term variations possibly caused by the PNS convection. To see the coherent lateral motion more clearly, we take time-averages of $Y_e$ and fluid velocity for the interval, $250<t<300$ms, the results of which are shown in the right panel of Fig.~\ref{g2Dvec_Yedis_V_contrast}. Note that in taking the average we adopt the PNS frame, i.e., the coordinate origin is shifted to the mass center of PNS. For comparison, we also calculate the average over the interval, $100<t<150$ms, and display it in the left panel.

The PNS convection has started before $100$ms; in fact several eddies can be recognized in the left panel. On the other hand, a large-scale fluid motion is clearly seen in the right panel, which is circulating counter-clockwise in the envelope of PNS at $15 \lesssim r \lesssim 25$km; it is particularly strong in the northern hemisphere; and there is a small region near the south pole, in which there is a small eddy circulating in the opposite sense. Such large-scale coherent circular motion may not be driven only by the PNS convection, since the lateral size exceeds the radial width of the convectively unstable region. We argue that this circular motion is triggered by the breaking of up-down symmetry owing to the PNS proper motion and is sustained by the combination of neutrino diffusion, asymmetric matter-pressure and PNS convection. Below we describe this mechanism in more detail.

Once the PNS core ($r<10$km) moves downwards, the matter distribution in the envelope of PNS starts to adjust itself, inducing a downflow near the north pole and a shear flow between the core and envelope of the PNS in the equatorial region. One may think that the velocity of these down- and shear-flows induced by the PNS proper motion should be of the same order as the velocity of the PNS proper motion ($\sim 10^{6} {\rm cm/s}$). The velocity of the circular motion observed in the simulation is $\sim 10^{8} {\rm cm/s}$, however. This may seem to indicate that the self-adjustment of the matter distribution in the PNS envelope can not be the origin of the circular motion. 

\begin{figure}
\vspace{15mm}
\epsscale{1.2}
\plotone{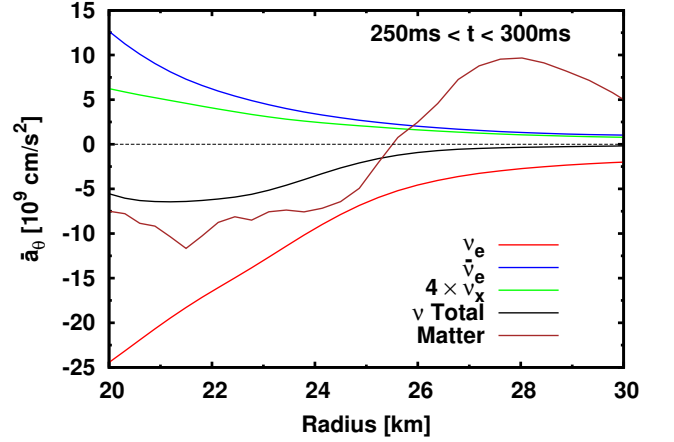}
\caption{Radial distributions of accelerations induced by the lateral diffusions of different species of neutrinos and matter-pressure. The time average is taken over the interval of $250{\rm ms}<t<300$ms (see Eq.~(\ref{eq:ac_theta_bydiffusion}) for the definition).
\label{graphNeutrinoPresGra}}
\end{figure}

\begin{figure}
\vspace{15mm}
\epsscale{1.2}
\plotone{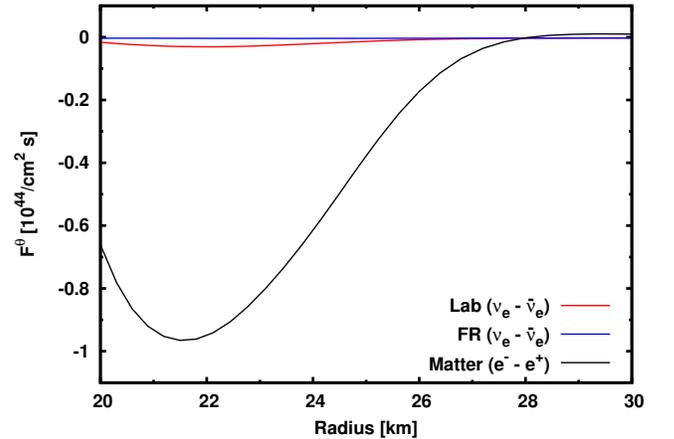}
\caption{Radial distributions of angle-average lateral number flux of $\nu_{\rm e} - \bar{\nu}_{\rm e}$ measured in the laboratory frame (red) and fluid-rest frame (blue). The counter part of electron/positron gas corresponds to the black line in the panel.
\label{graphLatefluxNeutrinoFRvsLabvsMatter}}
\end{figure}

This is not true, though. The most important role of the PNS proper motion is to break the up-down symmetry of the matter distribution around the PNS. On the north (south) side, the southward PNS-core motion expands (compresses) the PNS envelope. This then extends the low-$Y_e$ region downwards in the envelope of PNS on the north side but shrinks it on the opposite side. This asymmetry in the $Y_e$ distribution triggers the lateral diffusion of neutrinos. Due to the Fermi degeneracy of $\nu_e$, the diffusions of $\nu_{\rm e}$ and $\bar{\nu}_{\rm e}$ occur in the opposite directions (see also \citet{2018ApJ...854..136N}): $\nu_{\rm e}$ diffuses from higher- to lower-$Y_e$ regions (i.e., from south to north) and $\bar{\nu}_{\rm e}$ does in the opposite direction. Since neutrino diffusions are accompanied by momentum transfer to matter, they may account for the lateral circular motion.

We estimate the acceleration by the momentum feedback from neutrino to matter and matter-pressure gradient as
\begin{eqnarray}
&& \bar{a}_{\theta(i)}(r) \equiv - \left< \frac{ \frac{\partial}{\partial{\theta}} \bar{P}_{i}(r, \theta) }{r \bar{\rho}(r, \theta) } \right> \nonumber \\
&& \sim  - \frac{  \biggl( \bar{P}_{i}(r, \theta= \frac{3 \pi}{4})  -  \bar{P}_{i}(r, \theta= \frac{\pi}{4}) \biggr) }
{  (0.5 \pi) r \bar{\rho}_{\rm ave}(r)  }, \label{eq:ac_theta_bydiffusion}
\end{eqnarray}
where the symbols $~\bar{}~$ and $\left< ~ \right>$ denote the time and lateral averages, respectively; the subscript "$i=(\nu, m)$" specifies the neutrino or matter contribution; $P_{\nu}$ stands for the neutrino-pressure, which is approximately evaluated as $E_{\nu}/3$ in the diffusion regime, and $P_{m}$ denotes the matter-pressure; $\rho_{\rm ave}$ is the angular average of baryon mass density. In Fig.~\ref{graphNeutrinoPresGra}, we show their radial distributions for the time interval, $250{\rm ms}<t<300$ms.

As is clear in the figure, $\nu_{\rm e}$ dominates the acceleration and pushes the matter upward, as expected. We also find that the matter-pressure gradient contributes to the lateral acceleration on the same level. It is mainly driven by $Y_e$ asymmetry: the electron/positron pressure is higher in the southern hemisphere, where $Y_e$ is larger, and the matter is accelerated upward (but see below for more quantitative arguments). As displayed in Fig.~\ref{g2Dvec_Yedis_V_contrast}, the total acceleration is strong at $20<r<25$km, which coincides with the region, where we see the fastest fluid motion from south to north (see the right panel of the same figure). Since the flow is confined in a closed region, a counter-flow occurs deeper ($15 \lesssim r \lesssim 20$km) in the PNS envelope (see also the discussions below). The matter velocity ($v^{\theta}$) produced by the neutrino diffusion and matter-pressure gradient can be estimated more quantitatively as
\begin{eqnarray}
&&v^{\theta} \sim (2 \hspace{0.5mm} \bar{a}_{\theta} \hspace{0.5mm} L)^{0.5} \nonumber \\
&& \sim 2 \times 10^{8} \hspace{0.5mm} [{\rm cm/s}] \hspace{0.5mm} \left( \frac{\bar{a}_{\theta}}{10^{10} {\rm cm/s^2}} \right)^{0.5} \hspace{0.5mm} \left( \frac{L}{20 {\rm km}} \right)^{0.5}, \label{eq:v_theta_ndif}
\end{eqnarray}
where $L$ is the spatial scale of the lateral motion. This is consistent indeed with the velocity observed in the simulation. We thus conclude that both the neutrino diffusion and asymmetric matter-pressure gradient play a major role in the generation of the lateral circular motion.

To see the lateral transport of lepton number more in detail, we display the radial distributions of angle-averaged lateral lepton-number fluxes carried by neutrinos ($F^{\theta}_{\nu_{\rm e}} - F^{\theta}_{\bar{\nu}_{\rm e}}$, where $F^{\theta}$ denotes the lateral number flux of neutrinos) and by electron/positron in Fig.~\ref{graphLatefluxNeutrinoFRvsLabvsMatter}. As shown there, the latter is the main contributor to the lepton number transport; indeed, the neutrino contribution is just a few percent. This is simply because the number density of electron/positron is much larger than that of neutrino. The neutrino diffusion in the fluid rest frame is even subtler although it does exist (see the blue line in Fig.~\ref{graphLatefluxNeutrinoFRvsLabvsMatter}). It should be noted, however, that the momentum feedback from neutrinos is still comparable to the momentum imparted by the matter-pressure gradient, as discussed above. This happens because the lateral matter-pressure gradient is partially compensated by the temperature asymmetry: the temperature in the northern hemisphere is $\sim 1 \%$ higher than in the southern hemisphere. Although the temperature asymmetry is much smaller than that of $Y_e$, it still reduces the total asymmetry in the matter pressure substantially, since the baryon and photon pressure are dominant in the total pressure.

It is also important that $Y_e$ decreases via deleptonization as the matter moves toward the north pole. The $Y_e$-depleted matter then submerges in the PNS envelope near the north-pole. When it hits the PNS core, it is deflected to flow southward along the surface of the PNS core, increasing $Y_e$ this time by $\nu_{\rm e}$ absorptions. The matter emerges again to the PNS surface by the local PNS convection near the south pole. Although the strong downflow induced by the dislocation of PNS tends to suppress the convection around the north-pole, the convection activity continues to occur in the vicinity of the south-pole and supplies the $Y_e$-rich material to the PNS envelope there. This closes the cycle and creates the coherent lateral circular motion with the $Y_e$ asymmetry sustained. Then the circulation will continue until the PNS convection shuts off or the neutrinos in PNS are dried up, both of which may take more than a few seconds. If that is really the case, the PNS may be accelerated to more than a hundred km/s, the typical proper motion velocity of pulsars.

Finally, we give some comments on other works on the asymmetric $Y_e$ distributions. \citet{2018arXiv180910150G} found a circular motion in the envelope of PNS in their 3D simulations, which produced the LESA phenomenon. They speculated that the thermal instability of spherical shells discussed by \citet{1961hhs..book.....C} may be responsible for the sustained asymmetric $Y_e$ distribution. It seems, however, that the circular motion found in their study was not as coherent as what we observed in our study; indeed, they look more like a superposition of some smaller eddies instead (see Figure~6 in \citet{2018arXiv180910150G}). \citet{2018arXiv181205738P} also observed LESA in their simulations and suggested that LESA is in effect a manifestation of PNS convections under a different guise. It should be noted that they did not find any coherent circular motions in the envelope of PNS. Importantly, both simulations sphericalized the matter distribution inside PNS by hand and may have artificially suppressed the proper motion of PNS, which we think is the trigger of the coherent circular motion by breaking the up-down symmetry of the matter distribution around PNS.

\section{Summary and discussion} \label{sec:summary}
For decades, asymmetric neutrino emissions have been regarded as a minor player in the generation of NS kick unless some extreme conditions or unknown physics are considered. In our CCSN simulation, however, we find $\sim 10 \%$ levels of asymmetry in the energy flux for $\nu_{\rm e}$ and $\bar{\nu}_{\rm e}$, and $\sim 1 \%$ asymmetry for $\nu_x$. The emergence of the coherent asymmetric configurations coincides with the onset of the non-spherical shock expansion and they are sustained at least up to the end of the simulation. Interestingly, they play a non-negligible role in the acceleration of PNS. The asymmetry of $\nu_{\rm e}$ and $\bar{\nu}_{\rm e}$ is attributed to that of the $Y_e$ distribution in the envelope of PNS ($10 \lesssim r \lesssim 25$km). The self-adjustment of the matter distribution accompanied by the neutrino diffusion there generates a sustained lateral circulation, which in turn works to maintain the asymmetry in the $Y_e$ distribution. It is important to emphasize that all of these dynamics can be captured only with the consistent treatment of neutrino transport and PNS proper motions. Incidentally, \citet{2014ApJ...792...96T, 2017ApJ...837...84J} speculated that LESA may also contribute to the NS kick but did not demonstrate them quantitatively in their works.

Finally, we put several remarks. As is well known, the imposed axisymmetry artificially enhances the directionality in fluid dynamics such as shock expansions and PNS proper motions. The acceleration of PNS may hence be overestimated in this study. Nevertheless the property of asymmetric neutrino emissions ($\bar{\nu}_{\rm e}$ and $\nu_x$ emissions are enhanced in the hemisphere of stronger shock expansion, while $\nu_{\rm e}$ emissions are higher in the opposite hemisphere, into which the PNS is accelerated) will remain in 3D although it should be confirmed in the future study. A longer simulation is definitely required in order to assess how the PNS proper motion found in this paper affects the hydrodynamic process that will occur later and is supposed to be the most promising mechanism for the NS kick. It should be emphasized that they are operational in different phases. Indeed, the current simulation is way too short to see the PNS acceleration by the gravitational tugboat mechanism \citep{2013A&A...552A.126W,2017ApJ...837...84J}, for example. What is important regardless is that in some (but not all) cases, the PNS may start to move much earlier on than previously thought. Last but not least, the asymmetry in neutrino emissions will affect the collective neutrino oscillation (Nagakura et al. 2019 in prep) and the nucleosynthesis \citep{2019arXiv190609553F}, both of which are currently being investigated in more details and will be published elsewhere.

\acknowledgments 
 We acknowledge Adam Burrows, David Radice and David Vartanyan for fruitful discussion. The numerical computations were performed on the supercomputers at K, at AICS, FX10 at Information Technology Center of Nagoya University. Large-scale storage of numerical data is supported by JLDG constructed over SINET4 of NII. H.N. was supported by Princeton University through DOE SciDAC4 Grant DE-SC0018297 (subaward 00009650). This work was also supported by Grant-in-Aid for the Scientific Research from the Ministry of Education, Culture, Sports, Science and Technology (MEXT), Japan (15K05093, 25870099, 26104006, 16H03986, 17H06357, 17H06365), HPCI Strategic Program of Japanese MEXT and K computer at the RIKEN (Project ID: hpci 160071, 160211, 170230, 170031, 170304, hp180179, hp180111, hp180239).
\bibliography{bibfile}

\begin{thebibliography}{}
\expandafter\ifx\csname natexlab\endcsname\relax\def\natexlab#1{#1}\fi
\providecommand{\url}[1]{\href{#1}{#1}}
\providecommand{\dodoi}[1]{doi:~\href{http://doi.org/#1}{\nolinkurl{#1}}}
\providecommand{\doeprint}[1]{\href{http://ascl.net/#1}{\nolinkurl{http://ascl.net/#1}}}
\providecommand{\doarXiv}[1]{\href{https://arxiv.org/abs/#1}{\nolinkurl{https://arxiv.org/abs/#1}}}

\bibitem[{{Arzoumanian} {et~al.}(2002){Arzoumanian}, {Chernoff}, \&
  {Cordes}}]{2002ApJ...568..289A}
{Arzoumanian}, Z., {Chernoff}, D.~F., \& {Cordes}, J.~M. 2002, \apj, 568, 289,
  \dodoi{10.1086/338805}

\bibitem[{{Bisnovatyi-Kogan}(1993)}]{1993A&AT....3..287B}
{Bisnovatyi-Kogan}, G.~S. 1993, Astronomical and Astrophysical Transactions, 3,
  287, \dodoi{10.1080/10556799308230566}

\bibitem[{{Buras} {et~al.}(2006){Buras}, {Janka}, {Rampp}, \&
  {Kifonidis}}]{2006A&A...457..281B}
{Buras}, R., {Janka}, H.-T., {Rampp}, M., \& {Kifonidis}, K. 2006, \aap, 457,
  281, \dodoi{10.1051/0004-6361:20054654}

\bibitem[{{Chandrasekhar}(1961)}]{1961hhs..book.....C}
{Chandrasekhar}, S. 1961, {Hydrodynamic and hydromagnetic stability}

\bibitem[{{Chatterjee} {et~al.}(2005){Chatterjee}, {Vlemmings}, {Brisken},
  {Lazio}, {Cordes}, {Goss}, {Thorsett}, {Fomalont}, {Lyne}, \&
  {Kramer}}]{2005ApJ...630L..61C}
{Chatterjee}, S., {Vlemmings}, W.~H.~T., {Brisken}, W.~F., {et~al.} 2005, \apj,
  630, L61, \dodoi{10.1086/491701}

\bibitem[{{Faucher-Gigu{\`e}re} \& {Kaspi}(2006)}]{2006ApJ...643..332F}
{Faucher-Gigu{\`e}re}, C.-A., \& {Kaspi}, V.~M. 2006, \apj, 643, 332,
  \dodoi{10.1086/501516}

\bibitem[{{Foglizzo} {et~al.}(2015){Foglizzo}, {Kazeroni}, {Guilet}, {Masset},
  {Gonz{\'a}lez}, {Krueger}, {Novak}, {Oertel}, {Margueron}, {Faure}, {Martin},
  {Blottiau}, {Peres}, \& {Durand}}]{2015PASA...32....9F}
{Foglizzo}, T., {Kazeroni}, R., {Guilet}, J., {et~al.} 2015, Publications of
  the Astronomical Society of Australia, 32, e009, \dodoi{10.1017/pasa.2015.9}

\bibitem[{{Fujimoto} \& {Nagakura}(2019)}]{2019arXiv190609553F}
{Fujimoto}, S.-i., \& {Nagakura}, H. 2019, arXiv e-prints, arXiv:1906.09553.
\newblock \doarXiv{1906.09553}

\bibitem[{{Fuller} {et~al.}(2003){Fuller}, {Kusenko}, {Mocioiu}, \&
  {Pascoli}}]{2003PhRvD..68j3002F}
{Fuller}, G.~M., {Kusenko}, A., {Mocioiu}, I., \& {Pascoli}, S. 2003, \prd, 68,
  103002, \dodoi{10.1103/PhysRevD.68.103002}

\bibitem[{{Furusawa} {et~al.}(2017){Furusawa}, {Togashi}, {Nagakura},
  {Sumiyoshi}, {Yamada}, {Suzuki}, \& {Takano}}]{2017JPhG...44i4001F}
{Furusawa}, S., {Togashi}, H., {Nagakura}, H., {et~al.} 2017, Journal of
  Physics G Nuclear Physics, 44, 094001, \dodoi{10.1088/1361-6471/aa7f35}

\bibitem[{{Gessner} \& {Janka}(2018)}]{2018ApJ...865...61G}
{Gessner}, A., \& {Janka}, H.-T. 2018, \apj, 865, 61,
  \dodoi{10.3847/1538-4357/aadbae}

\bibitem[{{Glas} {et~al.}(2018){Glas}, {Janka}, {Melson}, {Stockinger}, \&
  {Just}}]{2018arXiv180910150G}
{Glas}, R., {Janka}, H.~T., {Melson}, T., {Stockinger}, G., \& {Just}, O. 2018,
  arXiv e-prints, arXiv:1809.10150.
\newblock \doarXiv{1809.10150}

\bibitem[{{Hobbs} {et~al.}(2005){Hobbs}, {Lorimer}, {Lyne}, \&
  {Kramer}}]{2005MNRAS.360..974H}
{Hobbs}, G., {Lorimer}, D.~R., {Lyne}, A.~G., \& {Kramer}, M. 2005, \mnras,
  360, 974, \dodoi{10.1111/j.1365-2966.2005.09087.x}

\bibitem[{{Holland-Ashford} {et~al.}(2019){Holland-Ashford}, {Lopez}, \&
  {Auchettl}}]{2019arXiv190406357H}
{Holland-Ashford}, T., {Lopez}, L.~A., \& {Auchettl}, K. 2019, arXiv e-prints,
  arXiv:1904.06357.
\newblock \doarXiv{1904.06357}

\bibitem[{{Holland-Ashford} {et~al.}(2017){Holland-Ashford}, {Lopez},
  {Auchettl}, {Temim}, \& {Ramirez-Ruiz}}]{2017ApJ...844...84H}
{Holland-Ashford}, T., {Lopez}, L.~A., {Auchettl}, K., {Temim}, T., \&
  {Ramirez-Ruiz}, E. 2017, \apj, 844, 84, \dodoi{10.3847/1538-4357/aa7a5c}

\bibitem[{{Janka}(2017)}]{2017ApJ...837...84J}
{Janka}, H.-T. 2017, \apj, 837, 84, \dodoi{10.3847/1538-4357/aa618e}

\bibitem[{Janka {et~al.}(2016)Janka, Melson, \& Summa}]{Janka:2016fox}
Janka, H.~T., Melson, T., \& Summa, A. 2016, Ann. Rev. Nucl. Part. Sci., 66,
  341, \dodoi{10.1146/annurev-nucl-102115-044747}

\bibitem[{{Katsuda} {et~al.}(2018){Katsuda}, {Morii}, {Janka},
  {Wongwathanarat}, {Nakamura}, {Kotake}, {Mori}, {M{\"u}ller}, {Takiwaki},
  {Tanaka}, {Tominaga}, \& {Tsunemi}}]{2018ApJ...856...18K}
{Katsuda}, S., {Morii}, M., {Janka}, H.-T., {et~al.} 2018, \apj, 856, 18,
  \dodoi{10.3847/1538-4357/aab092}

\bibitem[{{Kusenko} {et~al.}(2008){Kusenko}, {Mandal}, \&
  {Mukherjee}}]{2008PhRvD..77l3009K}
{Kusenko}, A., {Mandal}, B.~P., \& {Mukherjee}, A. 2008, \prd, 77, 123009,
  \dodoi{10.1103/PhysRevD.77.123009}

\bibitem[{{Lai}(2001)}]{2001LNP...578..424L}
{Lai}, D. 2001, {Neutron Star Kicks and Asymmetric Supernovae}, ed.
  D.~{Blaschke}, N.~K. {Glendenning}, \& A.~{Sedrakian}, 424

\bibitem[{{Lyne} \& {Lorimer}(1994)}]{1994Natur.369..127L}
{Lyne}, A.~G., \& {Lorimer}, D.~R. 1994, \nat, 369, 127,
  \dodoi{10.1038/369127a0}

\bibitem[{{Mirizzi} {et~al.}(2016){Mirizzi}, {Tamborra}, {Janka}, {Saviano},
  {Scholberg}, {Bollig}, {H{\"u}depohl}, \&
  {Chakraborty}}]{2016NCimR..39....1M}
{Mirizzi}, A., {Tamborra}, I., {Janka}, H.~T., {et~al.} 2016, Nuovo Cimento
  Rivista Serie, 39, 1, \dodoi{10.1393/ncr/i2016-10120-8}

\bibitem[{{M{\"u}ller}(2016)}]{2016PASA...33...48M}
{M{\"u}ller}, B. 2016, Publications of the Astronomical Society of Australia,
  33, e048, \dodoi{10.1017/pasa.2016.40}

\bibitem[{{M{\"u}ller} {et~al.}(2012){M{\"u}ller}, {Janka}, \&
  {Marek}}]{2012ApJ...756...84M}
{M{\"u}ller}, B., {Janka}, H.-T., \& {Marek}, A. 2012, \apj, 756, 84,
  \dodoi{10.1088/0004-637X/756/1/84}

\bibitem[{{M{\"u}ller} {et~al.}(2018){M{\"u}ller}, {Tauris}, {Heger},
  {Banerjee}, {Qian}, {Powell}, {Chan}, {Gay}, \&
  {Langer}}]{2018arXiv181105483M}
{M{\"u}ller}, B., {Tauris}, T.~M., {Heger}, A., {et~al.} 2018, ArXiv e-prints.
\newblock \doarXiv{1811.05483}

\bibitem[{{Nagakura} {et~al.}(2019{\natexlab{a}}){Nagakura}, {Furusawa},
  {Togashi}, {Richers}, {Sumiyoshi}, \& {Yamada}}]{2019ApJS..240...38N}
{Nagakura}, H., {Furusawa}, S., {Togashi}, H., {et~al.} 2019{\natexlab{a}},
  \apjs, 240, 38, \dodoi{10.3847/1538-4365/aafac9}

\bibitem[{{Nagakura} {et~al.}(2017){Nagakura}, {Iwakami}, {Furusawa},
  {Sumiyoshi}, {Yamada}, {Matsufuru}, \& {Imakura}}]{2017ApJS..229...42N}
{Nagakura}, H., {Iwakami}, W., {Furusawa}, S., {et~al.} 2017, \apjs, 229, 42,
  \dodoi{10.3847/1538-4365/aa69ea}

\bibitem[{{Nagakura} {et~al.}(2014){Nagakura}, {Sumiyoshi}, \&
  {Yamada}}]{2014ApJS..214...16N}
{Nagakura}, H., {Sumiyoshi}, K., \& {Yamada}, S. 2014, \apjs, 214, 16,
  \dodoi{10.1088/0067-0049/214/2/16}

\bibitem[{{Nagakura} {et~al.}(2019{\natexlab{b}}){Nagakura}, {Sumiyoshi}, \&
  {Yamada}}]{2019arXiv190610143N}
---. 2019{\natexlab{b}}, arXiv e-prints, arXiv:1906.10143.
\newblock \doarXiv{1906.10143}

\bibitem[{{Nagakura} {et~al.}(2018){Nagakura}, {Iwakami}, {Furusawa}, {Okawa},
  {Harada}, {Sumiyoshi}, {Yamada}, {Matsufuru}, \&
  {Imakura}}]{2018ApJ...854..136N}
{Nagakura}, H., {Iwakami}, W., {Furusawa}, S., {et~al.} 2018, \apj, 854, 136,
  \dodoi{10.3847/1538-4357/aaac29}

\bibitem[{{Nordhaus} {et~al.}(2012){Nordhaus}, {Brandt}, {Burrows}, \&
  {Almgren}}]{2012MNRAS.423.1805N}
{Nordhaus}, J., {Brandt}, T.~D., {Burrows}, A., \& {Almgren}, A. 2012, \mnras,
  423, 1805, \dodoi{10.1111/j.1365-2966.2012.21002.x}

\bibitem[{{Nordhaus} {et~al.}(2010){Nordhaus}, {Brandt}, {Burrows}, {Livne}, \&
  {Ott}}]{2010PhRvD..82j3016N}
{Nordhaus}, J., {Brandt}, T.~D., {Burrows}, A., {Livne}, E., \& {Ott}, C.~D.
  2010, \prd, 82, 103016, \dodoi{10.1103/PhysRevD.82.103016}

\bibitem[{{Pajkos} {et~al.}(2019){Pajkos}, {Couch}, {Pan}, \&
  {O'Connor}}]{2019arXiv190109055P}
{Pajkos}, M.~A., {Couch}, S.~M., {Pan}, K.-C., \& {O'Connor}, E.~P. 2019, arXiv
  e-prints, arXiv:1901.09055.
\newblock \doarXiv{1901.09055}

\bibitem[{{Powell} \& {M{\"u}ller}(2018)}]{2018arXiv181205738P}
{Powell}, J., \& {M{\"u}ller}, B. 2018, arXiv e-prints, arXiv:1812.05738.
\newblock \doarXiv{1812.05738}

\bibitem[{{Richers} {et~al.}(2017){Richers}, {Nagakura}, {Ott}, {Dolence},
  {Sumiyoshi}, \& {Yamada}}]{2017ApJ...847..133R}
{Richers}, S., {Nagakura}, H., {Ott}, C.~D., {et~al.} 2017, \apj, 847, 133,
  \dodoi{10.3847/1538-4357/aa8bb2}

\bibitem[{{Sagert} \& {Schaffner-Bielich}(2008)}]{2008A&A...489..281S}
{Sagert}, I., \& {Schaffner-Bielich}, J. 2008, \aap, 489, 281,
  \dodoi{10.1051/0004-6361:20078530}

\bibitem[{{Scheck} {et~al.}(2006){Scheck}, {Kifonidis}, {Janka}, \&
  {M{\"u}ller}}]{2006A&A...457..963S}
{Scheck}, L., {Kifonidis}, K., {Janka}, H.~T., \& {M{\"u}ller}, E. 2006, \aap,
  457, 963, \dodoi{10.1051/0004-6361:20064855}

\bibitem[{{Scheck} {et~al.}(2004){Scheck}, {Plewa}, {Janka}, {Kifonidis}, \&
  {M{\"u}ller}}]{2004PhRvL..92a1103S}
{Scheck}, L., {Plewa}, T., {Janka}, H.~T., {Kifonidis}, K., \& {M{\"u}ller}, E.
  2004, \prl, 92, 011103, \dodoi{10.1103/PhysRevLett.92.011103}

\bibitem[{{Takiwaki} {et~al.}(2014){Takiwaki}, {Kotake}, \&
  {Suwa}}]{2014ApJ...786...83T}
{Takiwaki}, T., {Kotake}, K., \& {Suwa}, Y. 2014, \apj, 786, 83,
  \dodoi{10.1088/0004-637X/786/2/83}

\bibitem[{{Tamborra} {et~al.}(2014){Tamborra}, {Hanke}, {Janka}, {M{\"u}ller},
  {Raffelt}, \& {Marek}}]{2014ApJ...792...96T}
{Tamborra}, I., {Hanke}, F., {Janka}, H.-T., {et~al.} 2014, \apj, 792, 96,
  \dodoi{10.1088/0004-637X/792/2/96}

\bibitem[{{Togashi} \& {Takano}(2013)}]{2013NuPhA.902...53T}
{Togashi}, H., \& {Takano}, M. 2013, Nuclear Physics A, 902, 53,
  \dodoi{10.1016/j.nuclphysa.2013.02.014}

\bibitem[{{Winkler} \& {Petre}(2007)}]{2007ApJ...670..635W}
{Winkler}, P.~F., \& {Petre}, R. 2007, \apj, 670, 635, \dodoi{10.1086/522101}

\bibitem[{{Wongwathanarat} {et~al.}(2013){Wongwathanarat}, {Janka}, \&
  {M{\"u}ller}}]{2013A&A...552A.126W}
{Wongwathanarat}, A., {Janka}, H.-T., \& {M{\"u}ller}, E. 2013, \aap, 552,
  A126, \dodoi{10.1051/0004-6361/201220636}

\bibitem[{{Woosley} {et~al.}(2002){Woosley}, {Heger}, \&
  {Weaver}}]{2002RvMP...74.1015W}
{Woosley}, S.~E., {Heger}, A., \& {Weaver}, T.~A. 2002, Reviews of Modern
  Physics, 74, 1015, \dodoi{10.1103/RevModPhys.74.1015}

\end{thebibliography}

\end{document}